\documentclass[traditabstract]{aa} 
\usepackage{graphicx}
\usepackage{txfonts}
\usepackage{wasysym}
\usepackage{array}
\usepackage{natbib}
\bibpunct{(}{)}{;}{a}{}{,} 
\usepackage{ulem}

\newcommand{\mearth}{M_\oplus}
\newcommand{\rearth}{R_\oplus}
\newcommand{\msun}{M_\odot}

\newcommand{\mstar}{M_\star}
\newcommand{\rstar}{R_\star}

\newcommand{\prot}{P_\mathrm{rot}}
\newcommand{\arot}{A_\mathrm{rot}}
\newcommand{\rplanet}{R_\mathrm{p}}

\newcommand{\teff}{T_\mathrm{eff}}
\newcommand{\logg}{\log g}
\newcommand{\meh}{[\textrm{M}/\textrm{H}]}

\newcommand{\htwo}{\mathrm{H}_2}
\newcommand{\htwoo}{\mathrm{H}_2\mathrm{O}}
\newcommand{\chfour}{\mathrm{CH}_4}

\newcommand{\chisq}{\chi^2}
\newcommand{\chisqr}{\chi^2_\mathrm{r}}
\newcommand{\oneb}{\texttt{T1}/$B$}
\newcommand{\oner}{\texttt{T1}/$R$}
\newcommand{\twob}{\texttt{T2}/$B$}
\newcommand{\twor}{\texttt{T2}/$R$}

\usepackage{color}
\begin{document}

\title{Large Binocular Telescope view of the atmosphere of GJ1214b\thanks{Based 
on data acquired using the Large Binocular Telescope
(LBT). The LBT is an international collaboration among institutions 
in the United States, Italy, and Germany. LBT Corporation 
partners are the University of Arizona on behalf of the Arizona 
university system; Istituto Nazionale di Astrofisica, Italy;
LBT Beteiligungsgesellschaft, Germany, representing the Max-Planck 
Society, the Astrophysical Institute Potsdam, and Heidelberg 
University; the Ohio State University; and the Research
Corporation, on behalf of the University of Notre Dame, University 
of Minnesota and University of Virginia. Partly based on STELLA WiFSIP 
data (Strassmeier et al.~2004).}
}
\author{V.~Nascimbeni\inst{1,2}\thanks{email 
        address: \texttt{valerio.nascimbeni@unipd.it}} 
   \and M.~Mallonn\inst{3}
   \and G.~Scandariato\inst{4}  
   \and I.~Pagano\inst{4} 
   \and G.~Piotto\inst{1,2} 
   \and G.~Micela\inst{5}
   \and \\S.~Messina\inst{4}
   \and G.~Leto\inst{4}
   \and K.~G.~Strassmeier\inst{3}
   \and S.~Bisogni\inst{6}
   \and R.~Speziali\inst{7}
       }
\institute{INAF -- Osservatorio Astronomico di Padova, 
          vicolo dell'Osservatorio 5, 35122 Padova, Italy
      \and
           Dipartimento di Fisica e Astronomia, 
           Universit\`a degli Studi di Padova,
           Vicolo dell'Osservatorio 3, 35122 Padova, Italy
      \and
          Leibniz-Institut for Astrophysics,
          An der Sternwarte 16, 14482 Potsdam, Germany
      \and
          INAF -- Osservatorio Astrofisico di Catania, 
          via S. Sofia 78, 95123 Catania, Italy
      \and
          INAF -- Osservatorio Astronomico di Palermo, Piazza del Parlamento, 90134 Palermo, Italy
      \and
          INAF -- Osservatorio Astrofisico di Arcetri, largo E.~Fermi 5, 50125, Firenze, Italy
      \and 
      INAF -- Osservatorio Astronomico di Roma, via Frascati 33, 00040 Monte Porzio Catone (RM), Italy
          }

   \date{Submitted ---; accepted May 6, 2015; compiled \today}

\abstract{The atmospheric composition and vertical structure of the
  super-Earth GJ1214b has been a subject of debate since its discovery in 2009.
  Recent studies have indicated that high-altitude clouds
  might mask the lower layers. However, some data points that
were gathered at
  different times and facilities do not fit this picture,
  probably because of a combination of stellar activity and systematic
  errors.  We observed two transits of GJ1214b with the Large
Binocular Camera, the
  dual-channel camera at the Large Binocular Telescope. For the first
  time,  we simultaneously measured the relative planetary radius $k=\rplanet/\rstar$
  at blue and red optical wavelengths ($B+R$), thus
  constraining the  Rayleigh scattering on GJ1214b after
  correcting for stellar activity effects. To the same purpose, a
  long-term photometric follow-up of the host star was carried out
  with  WiFSIP at STELLA, revealing a rotational period that
is significantly
  longer than previously reported. Our new unbiased estimates of
  $k$ yield a flat transmission spectrum extending to  shorter
  wavelengths, thus confirming the cloudy atmosphere  scenario for
  GJ1214b.
}

   \authorrunning{Nascimbeni et al.}
   \titlerunning{An LBT view of the atmosphere of GJ1214b}
   \maketitle

\section{Introduction}\label{introduction}

The observational study of exoplanetary atmospheres is a very young
field of research that started in 2002 with the space-based detection of
sodium around a gaseous giant \citep{charbonneau2002} and was
followed a
few years later by the first such detection from the ground
\citep{redfield2008}.  In this context, transiting exoplanets offer a
great opportunity to characterize their atmospheres. During a transit
the amount of stellar light absorbed or scattered by the planetary
limb  (and therefore the ``effective'' planetary radius) is
wavelength-dependent and can be interpreted through models of
different chemical composition and physical structure
\citep{seager2010}.  While sharp atomic absorption lines such as
neutral Na and K are investigated by means of high-resolution
spectroscopy, low-resolution spectra and broad-band photometry can be
exploited to search for wide molecular bands in the near-infrared
(NIR), or for processes  affecting the continuum such as Rayleigh
scattering \citep{sing2011}.  The latter is more efficient in the
ultraviolet (UV) and blue optical regions due to its $\lambda^{-4}$
dependence. The positive or null detection of Rayleigh scattering can
greatly help in distinguishing  between families of atmospheric models
that otherwise would be  strongly degenerate, as shown by
\citet{howe2012}.
With the advent of more sophisticated instruments, astronomers are
focusing on characterizing smaller and smaller targets, now reaching a
class of planets with masses between 2 and 10 $\mearth$.  These
planets, dubbed super-Earths, lack any analog within our solar
system, and their possible composition is still a matter of debate.  The
missing parameters include the mean molecular weight $\mu$ of their
atmospheres, a key quantity that also constrains the inner structure
of these planets \citep{valencia2013}.

The first discovered member of the super-Earth class was GJ1214b
\citep{charbonneau2009}, a 2.8 $\rearth$, 6.3 $\mearth$ low-density
planet hosted by a M4.5V red dwarf \citep{anglada2013}. The atmosphere
of GJ1214b has been investigated with different techniques and
instruments by many authors (among them,  \citealt{desert2011} and
\citealt{bean2011}), sometimes presenting contradictory results. The
lack of consistency can  be explained with a combination of 1)
an incomplete treatment of correlated noise, also called \emph{\textup{red noise}}
\citep{pont2007}; 2) inhomogeneous assumptions (for example, on
orbital inclination: \citealt{southworth2008}); and 3) a high level of
stellar activity of the host GJ1214, which has been known for a long time
to be a spotted and flaring star \citep{berta2011}. More recent
works agree on detecting a flat and featureless transmission spectrum,
which rules out a cloudless, low-$\mu$ atmosphere as for a
solar-composition atmosphere \citep{berta2012,fraine2013,kreidberg2014}. Flat
spectra can be due both to high-altitude clouds masking the lower
layers, or to a high-$\mu$ atmosphere consisting, for instance, of methane or water
vapor.  The two model families are very difficult to distinguish
by standard techniques, although recently there is some consensus about
models based on clouds
\citep{morley2013,barstow2013,kreidberg2014}. Most of these works were
based  on NIR spectrophotometry.

Probing Rayleigh scattering could be useful to constrain $\mu$ in a
complementary way, as was done in our previous work about the hot
Uranus GJ3470b \citep{nascimbeni2013b}.  At the same time, it would
distinguish between whole classes of models that are based on mixtures of clouds
and hazes, as seen  for instance in Fig.~21 from \citet{howe2012}. The
best optical spectral region in which to carry out such a search is that
encompassing the  Johnson $U$ and $B$ bands, where the Rayleigh signal
is at its strongest.  Unfortunately, the blue flux of GJ1214, an
extremely red star with $B=16.4$, is too weak to allow
high-precision photometry with medium-sized telescopes.  The only
previous works that investigated this spectral region
\citep{demooij2012,demooij2013,narita2013b} delivered relatively
low-S/N light curves and large error bars on the transmission
spectrum. On the other hand, all those measurements were not
simultaneous, which gave rise to some doubts that activity effects  (which are
both wavelength- and time-dependent) might have biased the results.
As our knowledge on stellar activity progresses, it is becoming
clearer and clearer that simultaneous multiwavelength transit
observations, to be carried out with differential photometry or
spectrophotometry, are much more informative than sparse samplings
because at least the time-dependent  systematic component is zeroed
\citep{ballerini2012,oshagh2013}.

In this work, we present for the first time a simultaneous high-S/N
measurement of the relative radius $k=\rplanet/\rstar$ of GJ1214b in
the $B$ and $R$ bands, obtained through Large Binocular Camera
(LBC) dual-band photometry.
Stellar activity and rotation was monitored during the same season by
a long-term photometric  follow-up with WiFSIP at STELLA. This
paper is organized as follows. In Sect.~2 we describe how the
observations
were performed and how the light curves were extracted. In Sect.~3  we
estimate the period and amplitude of the  rotational modulation of
GJ1214b, exploiting this information to correct
our planetary radii derived from the LBC for activity effects.  In Sect.~4, we finally compare
our results with the literature by constructing and interpreting a
revised transmission spectrum.

\section{Observations and data reduction}\label{observations}

\subsection{LBC dual-channel photometry}

We observed two complete transits of GJ1214b during the nights of
March 29 and May 17, 2012 with the LBC camera mounted at the
double 8.4m Large Binocular Telescope (LBT). We refer to these two
events as \texttt{T1} and \texttt{T2} hereafter.  The LBC is a wide-field,
prime-focus imager consisting of two independent channels that
are optimized
for the optical UV/blue and red region, respectively
\citep{giallongo2008}.  In particular, the extremely high  blue
efficiency of this instrument, combined with its simultaneous
dual-channel capability, makes it unrivaled among other 8-10m class
facilities. The sky was clear and photometric on both nights.  We mounted
a Bessel $B$ and Bessel $R$ filter on the blue and red channel,
respectively. Exposure times were set to 30 s on both channels during
the first transit; during the second one, the exposure time on the
blue camera (where the flux of GJ1214 is much fainter) was  increased
to 60 s to reduce overheads and increase the duty cycle.  Hard
defocusing was applied to both cameras (up to $2''$ and $6''$ FWHM on
the $B$ and $R$ images, respectively) to avoid  saturation and
minimize pixel-to-pixel systematic errors. Only a  $5'\times 7'$ CCD
window from a single chip of the LBC mosaics  was read out to improve
speed. Nevertheless, the imaged  field of view (FOV) included about
fifteen reference stars with   a S/N of the same order of our
target.

\begin{figure*}
\centering \includegraphics[width=\columnwidth]{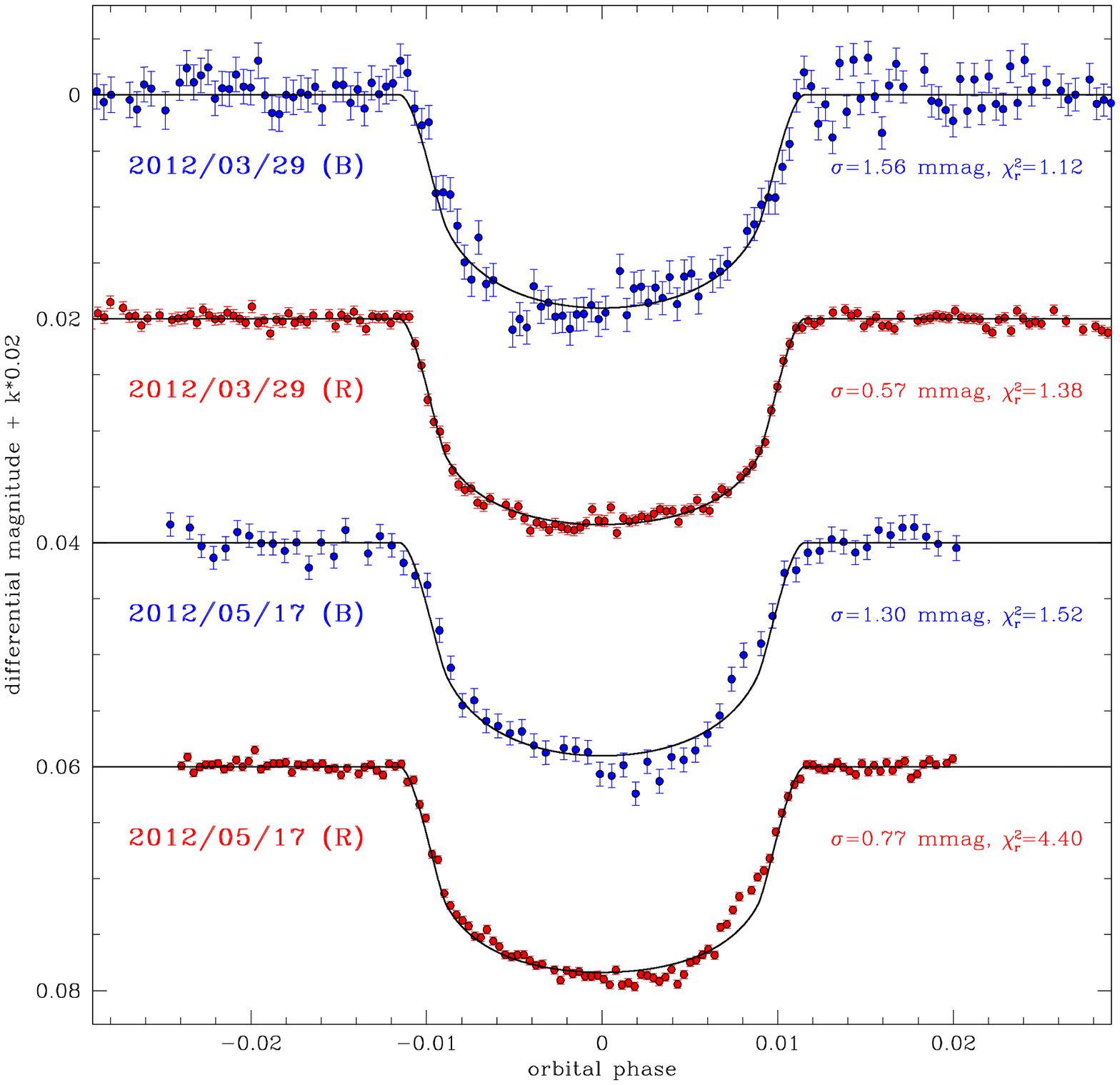}
\includegraphics[width=\columnwidth]{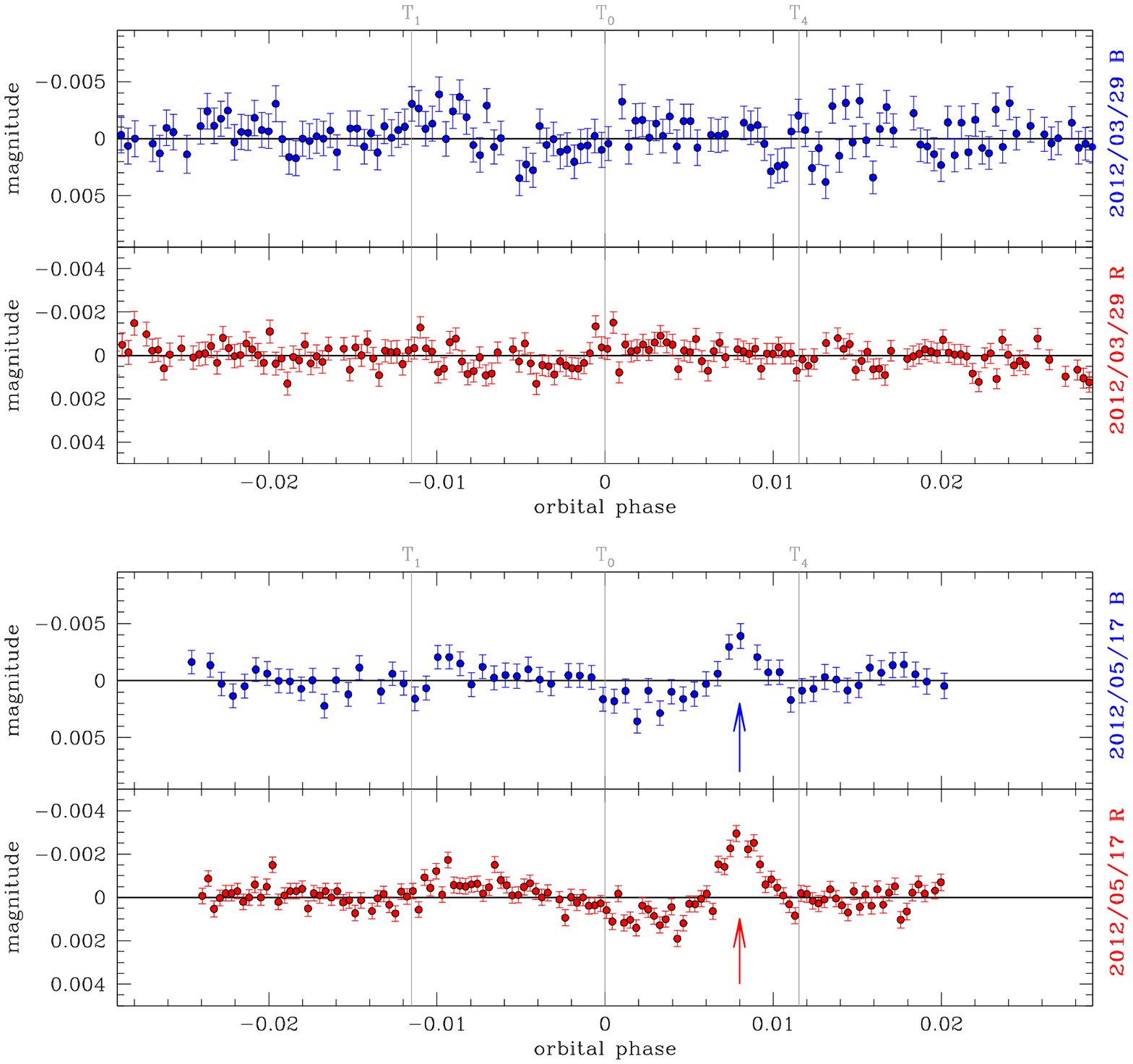}
\caption{\emph{Left panel:} light curves of GJ1214b  gathered through
  the blue channel of LBC in the Bessel $B$ band (blue points),  and
  through the red channel in Bessel $R$ (red points).  Data points are
  plotted with the original sampling cadence. The black line
  corresponds to the best-fit model adopting the parameters listed in
  Table \ref{results}. \emph{Right panel:} same as above, but showing
  the residuals from the best-fit model. The bump-like  feature at
  phase $0.070\lesssim\phi\lesssim0.092$ appeared during the second
  transit and is interpreted as a  starspot crossing; it is marked with an
  arrow in both channels.}
\label{lcs}
\end{figure*}

The raw images were corrected for bias and sky flat-field using
standard procedures. Stars drifted by about 50 pixels throughout the
series. This caused the target to cross a  CCD row that is larger
than the average pixel size; this is a result of a well-known manufacturing defect
first described by \citet{ak1999}.  This defect required a special
backward correction that was applied as prescribed by the same paper.
The light curves of GJ1214b were then extracted by running the STARSKY
pipeline, a code optimized to deliver high-precision differential
photometry over  defocused images, originally developed for the TASTE
project  \citep{nascimbeni2011a,nascimbeni2013a}. We refer
to our previous work on GJ3470b for further details on how STARSKY has
been fine-tuned on the LBC \citep{nascimbeni2013b}.  The four final light
curves, hereafter referred to as  \oneb, \oner, \twob, and \twor, are
shown in the left panel of Fig.~\ref{lcs}. Their  reduced $\chisq$,
evaluated\footnote{We define $\chisqr$ as the  $\chisq$ divided by the
  number of degrees of freedom, i.e., the number of data points minus
  the number of free parameters.} on the off-transit part  ranges
between $\chisqr=1.1$ and 1.5, meaning that our noise budget is
dominated by white-noise sources.  The dominant sources of noise are
expected to be mostly photon noise in the $B$ light curves and photon
noise plus scintillation in the $R$ light curves \citep{howell2006}.  A closer
inspection of the effect of correlated noise within the  overall noise
budget is presented in Sect. \ref{rednoise}.  The average
photometric error of the  \oner{} series is only 0.5 mmag over a 30 s
time scale,  which makes it one of the most accurate light curves of
GJ1214b ever taken  from a ground-based facility.

\subsection{STELLA photometric follow-up}

We monitored the host star GJ1214 in the observing seasons 2012 and
2013 to use its photometric variability as an indicator for the spot
coverage on the visible hemisphere at the time of the transit
observations. The program was performed with the 1.2m robotic twin
telescope STELLA, located in the Observatorio del Teide at Tenerife,
Canary Islands, and operated by the Leibniz-Institute for Astrophysics
Potsdam (AIP; \citealt{strassmeier2004}). We used the wide-field
imager WiFSIP, which consists of a 4k$\times$4k back-illuminated CCD
with a plate scale of $0.322''$/pixel and four read-out amplifiers. It
offers a field of view of $22'\times 22'$ on the sky
\citep{weber2012}. Between March 21 and October 10, 2012 we obtained
data on 110 nights; and between March 24 and October 4, 2013 we
observed on 81 nights. In 2012, we scheduled STELLA to observe GJ1214
with blocks of five exposures through a Johnson $V$ filter and an
exposure time of 150 s, followed by five exposures through a
Cousins $I$
filter with 60 s of integration time. To monitor the host star
activity at wavelengths more sensitive to starspot modulation, in 2013
we switched to filters Johnson $B$ (three exposures per block, 300 s of
exposure time each) and Johnson $V$ (three exposures, 150 s).  In total, we
gathered 197 images in $B$, 955 images in $V,$ and 739 in $I$. 

Bias and flat-field correction was applied by the STELLA data
reduction pipeline (for details see Granzer et al.~2015, in prep.).
The $I$ -band data showed fringe patterns whose shape varied mainly
with the CCD temperature.  In addition to changes in the sky OH
emission lines and the sky  continuum background, a second-order
effect in the fringe amplitude is due  to a change of opacity in the
CCD coating when the detector temperature changed  due to
low vacuum, for example.  We split all our $I$ -band measurements into bins of
different CCD temperatures and created a master fringe map for
each bin as the average of the individual images after they were
filtered
from  point-like sources and smoothed. The master fringe map of the
proper temperature was then scaled to match the  fringe amplitude of
each individual science frame and subtracted from it. This resulted in
a suppression of the fringe features to about 20\% of their original
value.  

Aperture photometry was performed with Source
Extractor\footnote{http://www.astromatic.net/software/sextractor}
\citep{bertin1996}.  We measured stellar fluxes with the Source
Extractor parameter \texttt{MAG\_AUTO}, which computes an elliptical
aperture according to the second-order moments of the object's light
distribution. This option provides the flexibility to account for very
different shapes of the PSF caused by different observing conditions
over the length of our follow-up program.  We applied stringent
selection criteria by rejecting data points  above a certain threshold
in airmass ($X=1.7$),  sky background ($2\,000$ ADUs), object FWHM and
elongation.  Furthermore, we rejected images where the target falls
onto the edge between different read-out amplifiers. The final sample
consisted of 128/436/425 images in $B$/$V$/$I$, respectively. To
perform  differential photometry, we constructed an artificial
comparison star  by summing the flux of several stable reference
stars. We tested several  selection criteria and found that a
combination of four nearby stars within $3'$ from GJ1214, imaged on
the same read-out amplifier, minimized the scatter in the final light
curve. The same comparison stars were employed for all three data sets
$B$, $V$, and $I$.

\begin{figure*}
\centering \includegraphics[width=1.8\columnwidth]{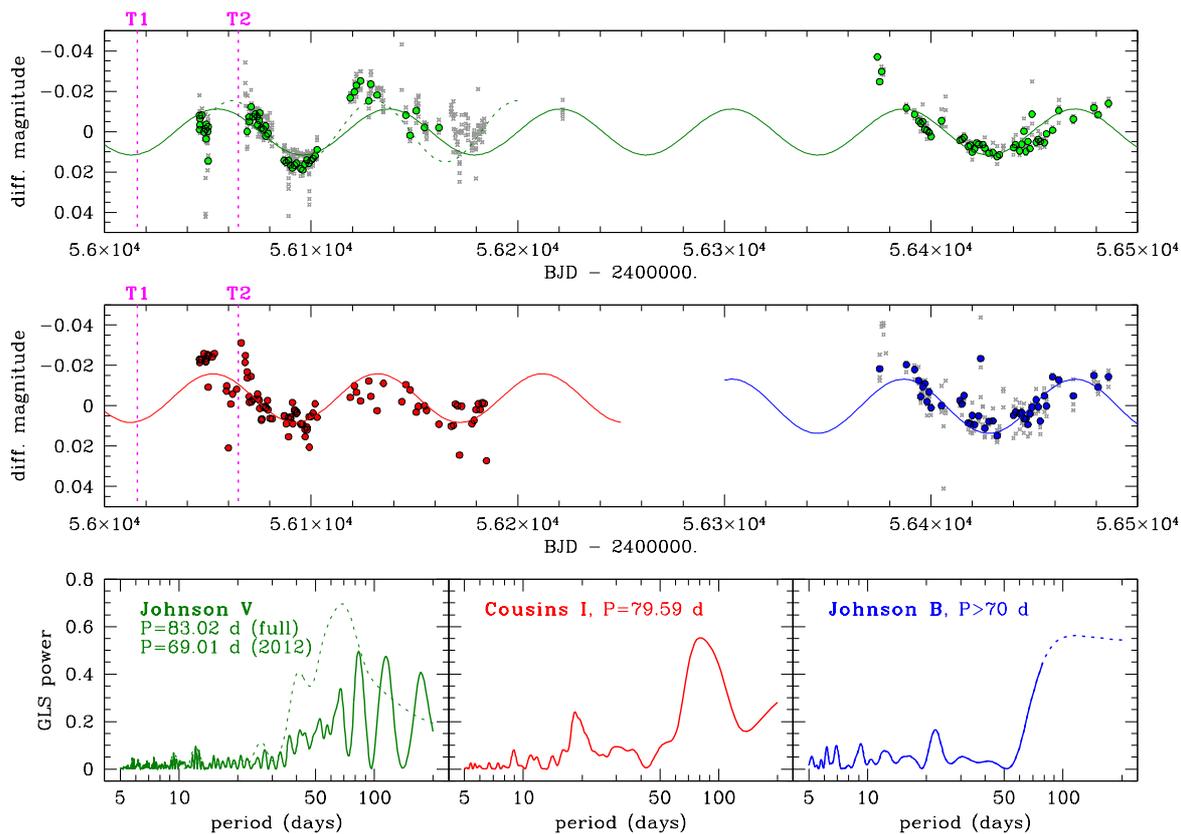}
\caption{\emph{Upper panel:} STELLA/WiFSIP differential  light curve
  of GJ1214b, gathered through a Bessel $V$  filter during the 2012
  and 2013 observing seasons. Unbinned points are shown in gray, daily
  averages are plotted with green points. Epochs corresponding to the
  two transits observed by LBC (\texttt{T1} and \texttt{T2}) are
  marked with vertical lines.  \emph{Middle panel:} same as above, but
  through a  Bessel $I$ filter in 2012 (red points) and Bessel $B$ in
  2013 (blue points). \emph{Lower panels:} GLS periodograms of the
   light curves, plotted with matching colors. The $V$
  periodogram has been evaluated on both the whole 2012/2013 data set
  (solid line) and only on the 2012 season (dashed line).  The
  best-fit period, or the lower limit to it, is reported for each
  periodogram.}
\label{stella}
\end{figure*}

The final light curves are shown in the two upper panels of
Fig.~\ref{stella} in their original cadence (gray dots). Light curves
were also binned by averaging all three or five points within a
single block, to obtain measurements that are much more robust
against outliers (green, red, and blue circles in Fig.~\ref{stella} for
the $V$, $I$, and $B$ data, respectively). From here on, our analysis
refers to the binned light  curves.

\section{Data analysis}\label{analysis}

\subsection{Starspot crossing event}\label{sce}

\begin{figure*}
\centering \includegraphics[width=\columnwidth]{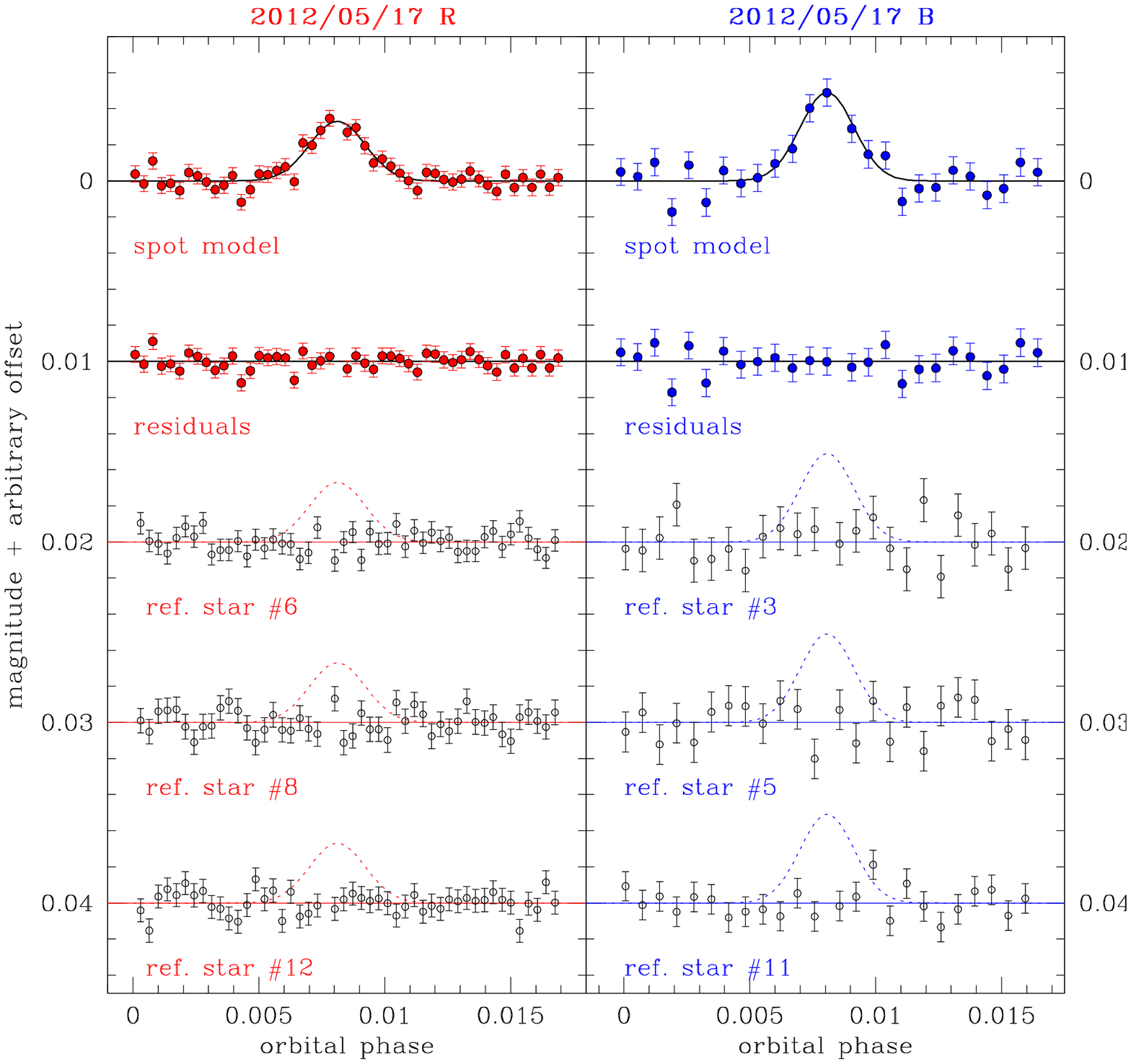}
\includegraphics[width=\columnwidth]{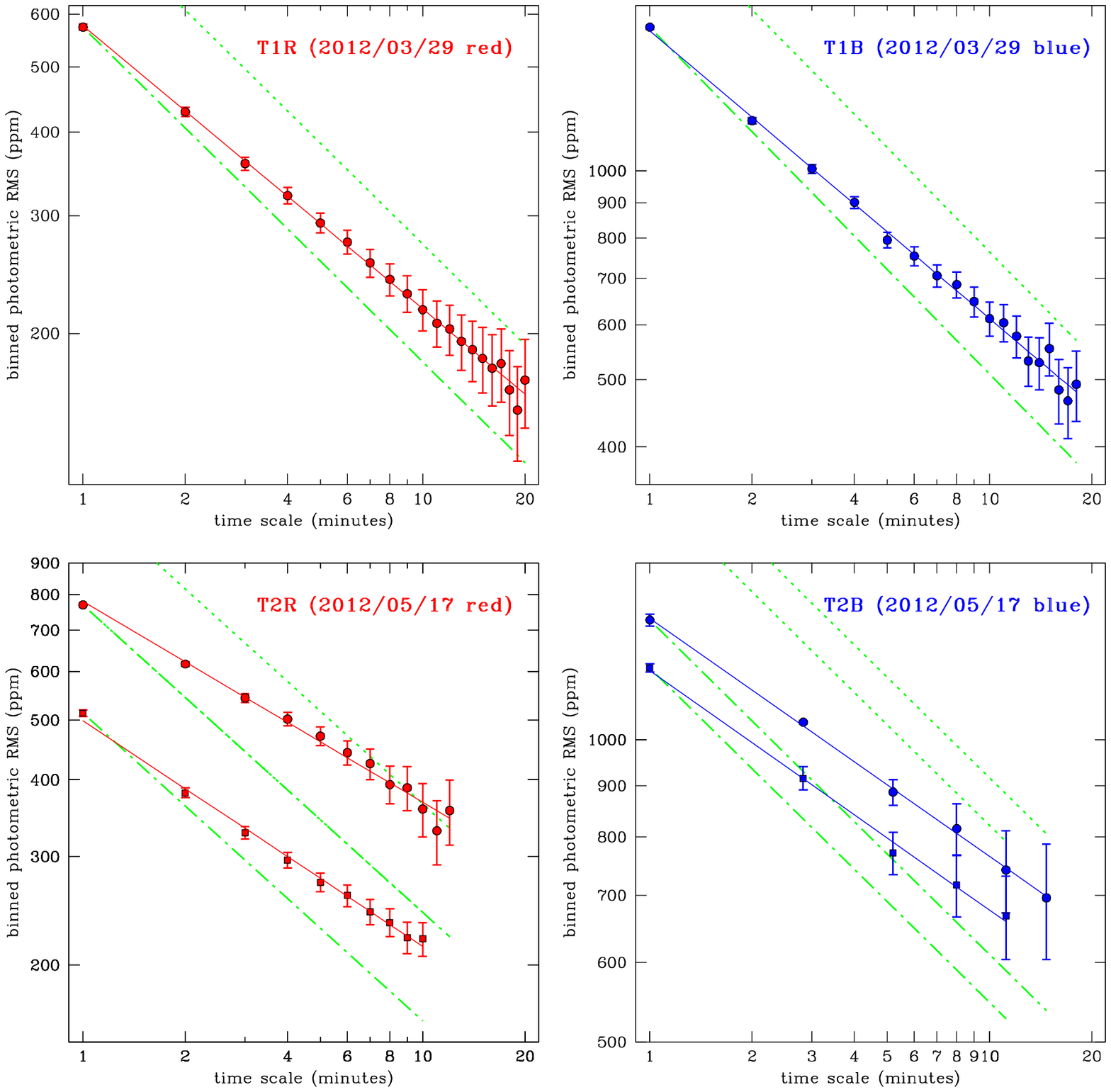}
\caption{\emph{Left panel:}  modeling of the \twor{} (left plots) and
  \twob{} (right plots) light-curve bumps. From the top: residual
  light curves with the  best-fit Gaussian model overplotted (black
  line; Table \ref{spotpars}), residuals from the Gaussian fit, and
  light curves of three reference stars with a red color and magnitude
  similar to GJ1214. Within the errors, the bump-like feature is  not
  seen in any other light curve.  \emph{Right panel:}  Correlated
  noise diagrams (photometric RMS of the binned time series as a
  function of bin size) for the GJ1214b LBC residual light curves  after the best-fit  transit model has been subtracted.  The lower
  sequence in the \twob{}, \twor{} plots is  evaluated after the
  starspot feature has been masked out. The green lines follow a
  Poisson-like $1/\sqrt{N}$ scaling law as extrapolated from the RMS
  of the unbinned curve (dotted-dashed line), and with a 50\%
  increase as reference (dotted line).}
\label{spot}
\end{figure*}

After a few preliminary fits of the LBC data, it became obvious that a
bump-like feature (marked with an arrow in the bottom right panel of
Fig.~\ref{lcs}) is clearly detectable in the residuals of both
channels, just before the egress of the second transit
(\texttt{T2}/$B$, \texttt{T2}/$R$).  We fitted each residual light
curve (i.e., after the transit model had been subtracted) with a
Gaussian function by leaving five free parameters:  its height over
the transit model $\delta$, its width $\sigma$, its center $t_0$, and
the coefficients $\Delta_0$, $\Delta_1$ of  a linear baseline
$\Delta_0+\Delta_1 t$.  The best-fit results, tabulated on
Table~\ref{spotpars} and plotted in the left panel of Fig.~\ref{spot},
were obtained through a nonlinear least-squares algorithm, while the
uncertainties were estimated though $10\,000$ iterations of a Monte
Carlo bootstrap.

Within the error bars, the duration and phase of the bump is
consistent in both light curves, while the maximum  amplitude is about
40\% larger in the $B$ band. This feature can be easily interpreted as
the crossing of an active region on the stellar surface. Indeed, the contrast between the unperturbed photosphere
and a colder starspot is higher at shorter wavelengths, giving rise to this typical color
signature.  We emphasize that the two LBC cameras are fully
independent, and so are the optical paths of the two LBT telescopes
when observing at the prime foci.  Furthermore, no similar feature has
been detected in either channel among  field stars with a red color
and similar magnitude to GJ1214 (Fig.~\ref{spot}, left panel, lowest
three curves).  We can therefore reject the hypothesis of an instrumental
or telluric origin of the bump. 

\begin{table}
\centering
\caption{Starspot parameters as fitted by a Gaussian model on the residual light
curves of the second transit (\twob{} and \twor{} light curves, Fig.~\ref{spot}).}
\begin{tabular}{lccl} \hline \hline
& \twob & \twor & unit \\ \hline
$\delta$ &  $4.9 \pm 0.8$          &     $3.3 \pm 0.3$           & mmag \\  
$t_0$ &  $8.07 \pm 0.16$  &     $8.12 \pm 0.09$   & orbital phase \textperthousand \\
&  $44.71 \pm 0.36\phantom{0}$       &     $44.83 \pm 0.20\phantom{0}$        & minutes from $T_1$ \\
&  $0.848 \pm 0.007$      &     $0.851 \pm 0.004$       & transit phase\\
$\sigma$ &  $1.07 \pm 0.19$&     $1.10 \pm 0.11$ & orbital phase \textperthousand \\
&  $2.44 \pm 0.43$        &     $2.50 \pm 0.24$         & minutes                  \\
FWHM  &  $5.7 \pm 1.0$          &     $5.9 \pm 0.6$           & minutes                  \\ \hline
\end{tabular}
\tablefoot{The transit phase is calculated by assuming the total 
duration of the eclipse ($T_1$-$T_4$, \citealt{kipping2010b}), in our case 52.4 
minutes. The FWHM is calculated from the Gaussian $\sigma$ as $2\sqrt{2\ln 2}\sigma$.}
\label{spotpars}
\end{table}
 
A few authors already reported tentative detections of starspots
occulted by GJ1214b
\citep{berta2011,carter2011,kundurthy2011,narita2013b}, but this is
the first firm detection based on multichannel photometry. The size
of this active region is  probably of the same order as GJ1214b: the
approximate timescale of the crossing is $5.8\pm 0.5$ min, which is
compatible with  the time the planet takes to cross a point-like
feature on its host star (6.2 minutes).  We caution that since the
crossing took place very close ($\sim$2 min before)  to the third
contact of the transit, where the angle between the  normal to the
stellar surface and the line of sight is $\theta\simeq 65^\circ$, the
amount of  photosphere covered by the occulted spot could be up to
about twice the disk area of GJ1214b ($\sim$2\%), which is a lower
limit for the overall spot coverage.

The more obvious effect of a starspot-crossing event on the derived
transit parameters is to decrease the best-fit value of $k$, as well
as perturbing the other parameters in unpredictable ways
\citep{oshagh2013}. To avoid this, in the following analysis we
masked the bump from the \twob{} and  \twor{} light curves by assigning
zero weight to the affected points, at orbital phase
$0.070\lesssim\phi\lesssim0.092$.

By assuming that the size of the spot is the same as the projected
planetary disk, we can obtain a crude estimate of the temperature
difference $\Delta T$ between the spot and the  unspotted
photosphere. If one compares the heights $\delta$ of the $B$ and $R$
bumps (Table \ref{spotpars}) with the depth of the underlying transit
model at the central instant of the spot (14.0 mmag under the
out-of-transit level), it is straightforward to derive the spot color
relative to the average stellar color: $(B-R)_\mathrm{spot} -
(B-R)_\star \simeq 0.17$ mag. By assuming  $\teff{}_\star \simeq
3250$~K \citep{anglada2013} and modeling both fluxes as  black-body
radiation, we obtain $\Delta T \simeq 110$~K.  This agrees with
previous observational studies that measured  $\Delta T \lesssim
500$~K on early-to-mid M dwarf stars
\citep{berdyugina2005,barnes2011,jackson2013}.  Our $\Delta T$ guess
is of course a lower limit to the true $\Delta T$, as in
principle the spot could be smaller than the occulting planetary disk,
and as a consequence, $T_\mathrm{spot}$ could be cooler than our
estimate.

\subsection{Transit parameter extraction}
\label{TPE}

The LBC light curves have to be fitted with a transit model
to extract the parameters of interest, above all the relative radius
$k=\rplanet/\rstar$ , which is the wavelength-dependent quantity needed
to construct the transmission spectrum. We used the JKTEBOP
code\footnote{\textrm{http://www.astro.keele.ac.uk/$\sim$jkt/codes/jktebop.html}}
version 34 \citep{southworth2004} to perform this task, adopting a
quadratic law to model the stellar limb darkening (LD;
\citealt{claret2004}).  In principle, the free parameters of the fit
are the central transit  time $T_0$, the orbital period $P$,  the
fractional stellar and planetary radii $\rstar/a$ and $\rplanet/a$,
the orbital inclination $i$, the $\Delta_0$, $\Delta_1$ coefficients
of a linear baseline  $\Delta_0 + \Delta_1t$, and the linear and
quadratic LD terms $u_1$, $u_2$. Within this model, $k$ is a derived
parameter obtained by dividing $\rplanet/a$ by $\rstar/a$.

We did not fit for both LD terms,  following a common practice to
improve the robustness of the fit  \citep{southworth2008}. Instead,
$u_2$ was fixed on all light curves at its theoretical value for the
corresponding photometric band by interpolating the LD tables by
\citet{claret2012} computed for cool stars ($1500<\teff<4800$~K). We
adopted for GJ1214 the most recent values of  $\teff$, $\logg$ and
$\meh$ estimated by \citet{anglada2013}. We investigated the effect on
our fit of the linear term $u_1$ by setting it as a free parameter,
or by keeping it fixed at its theoretical value over independent
JKTEBOP runs.  GJ1214b does not show any transit time variation (TTV)
within about 10 s \citep{harpsoe2013}, therefore we fixed both $T_0$ and $P$
at their most recent and accurate values
\citep{kreidberg2014}. Furthermore, the orbital inclination $i$ is
well known to be strongly correlated with $k$
\citep{southworth2005}. To avoid biases when comparing our results
with the literature, we fixed the inclination and scaled semi-axis at
their best-fit value $i=88^\circ.94$, $a/\rstar=14.97$ found by
\citet{bean2010}, following what has been done by most other authors
(including \citealt{bean2011,desert2011,croll2011,
demooij2012,murgas2012,teske2013,narita2013b,fraine2013,demooij2013}). 


After these considerations, we had only three or four
fitted parameters for each light curve:  $k$, $\Delta_0$, $\Delta_1$
and, when left free, $u_1$. The other parameters were injected into
the fit as Gaussian priors, as done in \citet{nascimbeni2013b}; their
dispersions were set to the published errors for $P$, $T_0$ and $i$ by
\citet{kreidberg2014} and \citet{bean2010}, and as the errors
propagated on $u_1$ and $u_2$ by assuming the  published uncertainties
on  $\teff$, $\logg$ and $\meh$ by \citet{anglada2013}.  We chose to
derive the uncertainties over the best-fit parameters with a Monte
Carlo bootstrap algorithm (JKTEBOP task 7; \citealt{southworth2005})
because formal errors are known to be  underestimated due to
correlations between parameters and  residual systematic errors (the
so-called red noise; \citealt{pont2007}; a detailed analysis of it
follows in the next subsection). The results are plotted in Table
\ref{results},  the four individual and the two combined
light curves;  in the latter case, each pair of transits observed
through the same filter was fitted simultaneously. 

We note that when leaving $u_1$ free, the
best-fit value of the first epoch is consistent with the theoretical prediction. The
best-fit values of $k$ are also consistent with those obtained by
fitting  the light curves of the second transit in the matching
filter. For this  reason, and because the masking of the bump-like
feature does not appear to bias the $k$ measurement, the simultaneous
fitting of both transits appears justifiable.

\begin{table*}
\caption{Summary of the best-fit transit parameters of GJ1214b from LBC photometry.}
\centering\setlength{\extrarowheight}{3pt}
\begin{tabular}{lc|cccc|ccc} \hline \hline
transit    & band & $N_\mathrm{obs}$ & $\sigma$ [mmag] & $\chisq$ & $\chisqr$  & $u_1$           & $k=\rplanet/\rstar$& {\color{black}$k'=k+\Delta k_{\max}$}  \\ \hline
\texttt{T1}, 2012/03/29 & $B$ & 185  & 1.55 & 197.3 & 1.11 & $0.60\pm 0.07$  & $0.1158\pm0.0013$  & {\color{black} --- }  \\
\texttt{T1}, 2012/03/29 & $B$ & 185  & 1.56 & 200.2 & 1.12 & 0.5259$^\dagger$& $0.1170\pm0.0009$  & {\color{black} $0.1159\pm0.0009$ }  \\ 
\texttt{T1}, 2012/03/29 & $R$ & 207  & 0.57 & 282.2 & 1.39 & $0.42\pm 0.02$  & $0.1170\pm0.0004$  & {\color{black} $0.1159\pm0.0004$ }  \\ \hline
\texttt{T2}, 2012/05/17 & $B$ & 64   & 1.32 & \phantom{0}64.0 & 1.05 & 0.5259$^\dagger$& $0.1186\pm0.0010$  & {\color{black} --- }  \\
\texttt{T2}, 2012/05/17 & $R$ & 123  & 0.76 & 266.4 & 2.22 & 0.4326$^\dagger$& $0.1178\pm0.0004$  & {\color{black} --- }  \\ \hline
combined                & $B$ & 249  & 1.50 & 273.1 & 1.11 & 0.5259$^\dagger$& $0.1176\pm0.0007$  & {\color{black} $0.1165\pm0.0007$ }  \\
combined                & $R$ & 330  & 0.65 & 575.5 & 1.76 & 0.4326$^\dagger$& $0.1175\pm0.0003$  & {\color{black} $0.1164\pm0.0003$ }  \\ 
combined, RP            & $B$ & 249  & 1.50 & 273.1 & 1.11 & 0.5259$^\dagger$& $0.1176\pm0.0009$  & {\color{black} $0.1165\pm0.0009$ }  \\
combined, RP            & $R$ & 330  & 0.65 & 575.5 & 1.76 & 0.4326$^\dagger$& $0.1175\pm0.0004$  & {\color{black} $0.1164\pm0.0004$ }  \\ \hline
\end{tabular}
\tablefoot{The columns list: the ID code and evening date of the transit (or combined when the two 
transits were fitted together), the photometric channel (Bessel $B$ or $R$), the number of
data points analyzed, the RMS $\sigma$ of the residuals from the best-fit model, the $\chisq$ and reduced $\chisq$ of 
the residuals, the linear LD coefficient $u_1$ (marked with $\dagger$ when it is input as a 
Gaussian prior around its theoretical value from \citealt{claret2012}), the relative radius $k$ 
of the best-fit model, and the relative radius $k'$ after the correction for unocculted starspots
has been applied (\citealt{ballerini2012}; see Sect.~\ref{kmax} for details). The reported $\chisq$ and $\chisqr$
are evaluated from the nominal error bars once the bump-like feature has been removed (see Sect.~\ref{sce}). 
The RP error analysis is based on a residual-permutation algorithm (see Sect.~\ref{rednoise} for details).
}
\label{results}
\end{table*}


\subsection{Correlated noise analysis}
\label{rednoise}

After  the transits were properly modeled, we
investigated whether and to which extent the residual light curve
(i.e., after the transit best-fit model has been subtracted) is
affected by correlated noise. To this purpose, a commonly used tool is
the so-called correlated noise diagram  (CND), where the input
data are binned on different timescales and the resulting
photometric RMS of the binned time series is plotted as a function of
bin size \citep{pont2007,carter2009}.  We plotted CNDs for all our
four light curves (\oneb, \oner, \twob, and \twor) in the right panel of
Fig.~\ref{spot}. For the \twob{} and \twor{} transits, we  separately 
plotted the CNDs for the full light curve (upper sequence) and
after masking the starspot feature (lower sequence). The CNDs have to
be compared with a Poissonian $1/\sqrt{N}$ scaling law (where $N$ is
the number of points within each time bin), which is expected to
predict the RMS after the averaging process in the  ideal case when
only white noise contributes (dashed-dotted green line on
Fig.~\ref{spot}). More realistically, where correlated noise is not
negligible, we expect to see an increase of the RMS toward the longest
timescales.  This is clearly visible in our case.
For the \oner{} and \oneb{} light curves, the RMS increase is
always within  $\sim$20\% of the overall budget even at the longest
available timescales  (bin size $\sim$20 min, i.e., the maximum
allowed by our requirement of  having a set of at least ten bins 
to evaluate a meaningful RMS). This proves that the effect of
red noise on the first transit is marginal.  On the other hand, the
\twor{} and \twob{} light curves show more red noise,
which boosts the binned RMS by $\sim$50\% even if the  starspot
bump is not taken into account. This is also confirmed by
the best-fit linear slopes, which are steeper for the first transit:
$\sigma(\oner)\propto N^{-0.35}$, $\sigma(\oneb)\propto N^{-0.33}$, 
with respect to the second one:
$\sigma(\twor)\propto N^{-0.23}$, $\sigma(\twob)\propto N^{-0.22}$
(the latter values are calculated after masking the spot). 

\begin{figure*}
\centering \includegraphics[width=1.8\columnwidth]{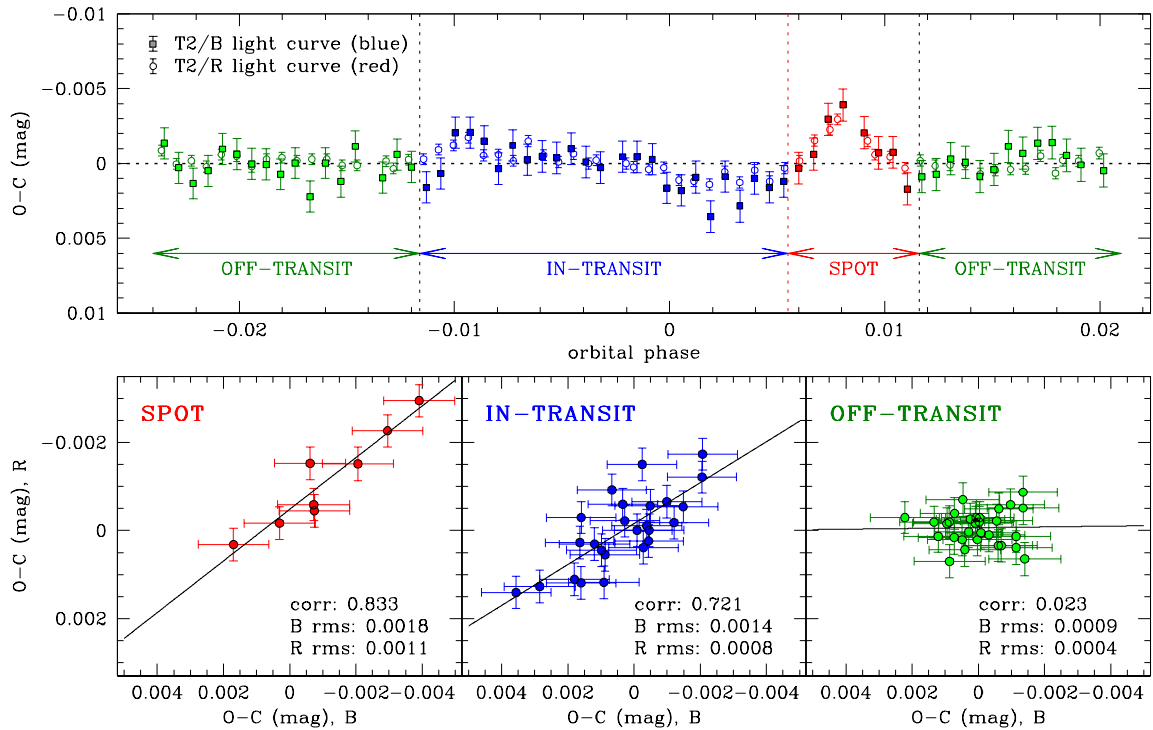}
\caption{Statistical correlation between the blue (\twob{}) and 
red (\twor{}) LBC light curve of the second transit.
\emph{Upper panel:} Observed minus calculated ($O-C$) residuals
from the best-fit transit model as a function of orbital phase for the
\twob{} (square points) and the \twor{} (round points) time series. 
Data are color-coded according to whether they are labeled as off-transit, in-transit
(the spot feature being excluded) and spot only. Only points whose 
mid-exposure time can be matched within 30~s to a frame of the other channel 
are considered here. 
\emph{Lower panels:} Correlation plots between residuals from \twob{} and 
\twor{} light curves for the three samples defined above. The best-fit
straight line calculated by ordinary least-squares is overplotted. The
Spearman rank correlation coefficient is tabulated for each subset, along with the
blue and red RMS.}
\label{corr}
\end{figure*}

Although the $\texttt{T2}$ light curve is still
dominated by photon noise at the 10-15~min timescale, the question is
whether this additional noise is of instrumental,
atmospheric, or astrophysical origin. A closer inspection of the
\twor{} vs. \twob{} residual curves (lower right panel of
Fig.~\ref{lcs}) reveals some correlation between the two channels, but
only during the transit.  
To support this hypothesis, we selected three subsets of
data points covering the off-transit part, the in-transit 
(spot excluded), and the spot feature itself (Fig.~\ref{corr}, upper panel). 
Points were then paired by matching mid-exposure times that differ by 
less than 30~s between the two channels. By comparing the $B$ and $R$ 
residuals of the spot and in-transit series, a strong positive
correlation is found (Spearman rank correlation coefficient: $\varrho = 0.833$ and 0.721,
respectively), while the off-transit points are essentially
uncorrelated ($\varrho = 0.023$). Interestingly, if we fit a straight
line to the first two subsets (Fig.~\ref{corr}, lower panels), even the
slope of the correlation is similar (0.58 vs.~0.46; i.e., the $B$ 
amplitude in on average about twice the $R$ one).
As a similar behavior is not seen in any
reference star, we conclude that most of  the measured red noise is
due to the crossing of additional inhomogeneities on the  GJ1214
spotted surface, which was caught here in a relatively active phase.
Additional evidence of such an increase of activity is exposed in the
next subsection.

To check whether and to which extent the correlated noise
content of our data might bias our derived transit parameters
(Table~\ref{results}), we performed an additional error analysis based
on a residual-permutation (RP) or prayer bead algorithm, which
is based on the cyclic permutation of the residuals
\citep{southworth2008}. This approach preserves the temporal structure
of the correlated noise and  propagates it through the whole light
curve, leading to more realistic error estimates.  We ran the
algorithm on our combined data sets, setting JKTEBOP with the same
input parameters described in the previous subsection. The resulting
best-fit values are $k=0.1176\pm0.0009$ ($B$) and $k=0.1175\pm0.0004$
($R$), which are fully consistent with the previous ones but with a
slightly larger error bar. Following a conservative choice, we 
adopted the RP best-fit values to build the transmission spectrum of
GJ1214b.


\subsection{Stellar activity and rotation}
\label{SAR}
All our three $B$, $V$, $I$ STELLA light curves of GJ1214
(Fig.~\ref{stella})  show clear signs of periodic variability at
similar timescales. The phase of that signal is coherent among
different filters ($V$ vs.~$I$ in 2012 observing  season, $B$ vs.~$V$
in 2013), while the amplitude is larger at shorter wavelengths.  This
signature is easily interpreted as rotational modulation, 
meaning that it is the result of an unevenly spotted photosphere on a rotating
star. The photometric variability and flaring behavior of GJ1214 is
expected for a  M4.5V dwarf and was already known before our
study \citep{kundurthy2011}, but  the rotational period $\prot$ is not
firmly constrained so far.  \citet{berta2011} tentatively detected a
signal with an amplitude of $\arot=3.5\pm 0.7$ mmag and $\prot\simeq 53$
days from their MEarth $(i+z)$-band light curves gathered in 2010,
but the authors themselves cautioned that their other data sets  (Mearth
2008, 2009 seasons and FLWO 2010 $V$-band photometry) are not able to
confirm that signal. A more recent multicolor  follow-up by
\citet{narita2013b} carried out at the MITSuME 50  cm telescope in
2012 found $\prot = 44.3\pm 1.2$ d, with $\arot =21\pm4$, $5.6\pm
0.8$,  and $3.2\pm 0.4$ mmag  on the $g'$, $R_c$, and $I_c$ bands,
respectively. In this case, the authors  cautioned that the $g'$-band
detection is weak because of low S/N and that the limited baseline of
their observations prevented them from probing longer periods. More in
general, it is possible that the true $\prot$ of a star is an integer
multiple (2:1 or 3:1) of the detected photometric period  because there
are multiple groups of active regions at  different
stellar longitudes \citep{walkowicz2013}.

We analyzed our $B$, $V$, and $I$ light curves, binned over each
observing block of three or five repeated exposures, through a
generalized Lomb-Scargle periodogram (GLS; \citealt{zechmeister2009}).
The error bar of each binned point was estimated as the root mean
square (RMS) of the single measurements.  For the $V$ series, which
spans two observing seasons, we  analyzed both the whole series and
each season separately. This is useful because typical active regions
are expected to evolve over timescales ranging from weeks to months,
which means that the
phase coherence could be  broken or attenuated when folding very
long time series. The resulting periodograms are plotted in the lower
panel of Fig.~\ref{stella} and the corresponding  best-fit GLS
solutions in Table~\ref{periods}. We emphasize that the overall shape
of the light curves and the resulting periodograms are not influenced
by a  particular choice of reference stars, as confirmed by further
checks.

All three light curves show long-term ($\sim$80 d) periodicities
detected at high confidence. The $V$ whole-series periodogram shows an
excess of power between 60 and 100 d (with maximum power at $P\simeq
83$ d),  which turns into a single prominent peak at $P\simeq 69$ d
when only the  2012 season is taken into account (i.e., that with the
longest coverage).  The formal false-alarm probability (FAP) is
$10^{-13}$ and $8\cdot10^{-13}$, respectively. The $I$ series reveals
a signal at $P\simeq 79$ d with an even higher level of confidence
(FAP: $7\cdot 10^{-16}$).  Through both $V$ and $I$ filters, the
pseudo-sinusoidal shape of the signals  and their phase coherence are
clearly visible in the light curves.  
A curvature perfectly consistent with the  $V$ light curve gathered in
the same season is noticeable for the $B$ series. 
As  the smaller baseline of the $B$
data ($\sim$100 d, although only $\sim$60 d are densely sampled)
prevents us from detecting signals at $P\gtrsim 70$ d, we assumed $P\simeq
70$ as a lower limit. All STELLA data therefore converge toward a
rotational  period of $\prot\simeq 80$ d, or an integer multiple of
it.   Interestingly, \citet{berta2011} reported $\prot = 81$ d as the
most significant period in their MEarth 2008 data, which matches our
results very well, with an FAP of $10^{-4}$. On the other hand, the
$44$-d period reported by \citet{narita2013b} could be interpreted as
a  2:1 harmonic of our peak, caused by two groups of active regions
present at that time. 

\begin{table}
\centering
\caption{Most significant photometric periodicities of GJ1214 
detected in STELLA/WiFSIP light curves by the GLS periodogram.}
\begin{tabular}{ccccc} \hline \hline
filter & season & $P$ (d) & $A$ (mmag) & FAP \\ \hline
$V_J$ & 2012-2013 & $83.0\pm 0.8$ & $11.4\pm 1.1$ & $1\cdot 10^{-13}$ \\ 
$V_J$ & 2012      & $69.0\pm 2.0$ & $15.1\pm 1.3$ & $8\cdot 10^{-13}$ \\ 
$I_c$ & 2012      & $79.6\pm 2.5$ & $12.0\pm 1.2$ & $7\cdot 10^{-16}$ \\ 
$B_J$ & 2013      & $\gtrsim 70$  & $\gtrsim 15$  & --- \\ \hline
\end{tabular}
\tablefoot{The columns list: the bandpass and observing season
of the photometric series,
the period $P$ and amplitude $A$ of the GLS best-fit solution, and
the formal false-alarm probability (FAP). The $B_J$ solution is a
lower limit (see Sect.~\ref{SAR} for details).}
\label{periods}
\end{table}

The $V$, $I$ photometric amplitudes measured by our group during the
2012 campaign (i.e., when the LBC transits were gathered)  are
typically two to three times larger than those reported by
\citet{berta2011} and \citet{narita2013b}. This suggests a phase of
stronger  activity undergone by GJ1214 in 2012. This hypothesis is
also supported by the occulted starspots in our
\texttt{T2} transit, and in the August 12, 2012 light curve published
by \citet{narita2013b}.  {\citet{berdyugina2005} and
\citet{strassmeier2009} indicated that for an M4 star the typical
temperature  contrast and filling factor of starspots are $\Delta T
\lesssim 500$ K and  $\lesssim 20\%$, respectively.  This agrees
with our estimate of $\Delta T \gtrsim 110$~K based on the
$B-R$ color of the starspot occulted during the \texttt{T2} transit,
as shown in Sect.~\ref{sce}.  Using the \texttt{BT-Settl} atmospheric
model \citep{allard2011} to simulate such an M4 spotted photosphere
under the above assumptions for $\Delta T $ and filling factor,  we
can indeed consistently recover the variability  amplitude of the
STELLA photometric series.

\subsection{Correction for unocculted starspots}\label{kmax}

To construct a transmission spectrum of GJ1214b, the key
quantity to derive from our LBC transit light curves is the radius
ratio $k=\rplanet/\rstar$. This quantity is directly related with the
transit depth and can be affected by stellar activity in two
complementary ways. The first, most obvious one is by occulted
starspots,  which give rise to bumps in the light curve that
force the fitting procedure to underestimate $k$. Provided that the
S/N of the  timeseries is high enough, these bumps are easily
identified and masked  by assigning them zero weight, as we did on our
\texttt{T2} transit.

A second, subtler way to add a bias on $k$ is due to
\emph{unocculted} spots, because they decrease the mean surface
brightness of the visible photosphere, therefore leading to
overestimating $k$. This effect is color-dependent (being stronger at
shorter wavelengths, where the contrast between the unspotted
photosphere and the cooler spots  is increased) and requires a
correction that takes into account the   spot coverage of the star
at the time of observation.  We followed the approach of
\citet{ballerini2012} to compute that correction and refer 
to our previous work on GJ3470b for further details
\citep{nascimbeni2013b}.  We neglected the contribution from faculae, as
no typical signatures of facular occultations by GJ1214b are  recorded
either in our data or in the literature light curves.  {Below, 
  we focus on investigating the impact of activity on
  $k$ during the first LBC epoch (\texttt{T1}),  where we can
  reasonably assume that there are no occulted starspots (case 2
  above).  The starspot-crossing event on \texttt{T2} and the possible
  presence of additional  minor active regions (see Sect.
  \ref{rednoise}) instead complicate the modeling, because a mixed
  scenario with both occulted and unocculted starspots cannot be
  approximated by either of the simplified cases above. 

At the time of \texttt{T1}  transit, the star appeared  $\Delta V
\simeq \Delta I \simeq  0.023$  mag fainter than the
maximum luminosity as it is modeled on the $V$/2012 and $I$/2012 light
curves by the best-fit  sinusoids plotted in Fig.~\ref{stella}.
Assuming that the net observed stellar flux is the sum of the quiet
and spotted photosphere, weighted by the corresponding coverage factor
over the visible hemisphere \citep{afram2015,scandariato2014}, this
translates into an estimated  spot coverage of $\sim$0.020 at the
epoch of observation.   Following \citet{ballerini2012}, adopting  $\Delta
T = 500$ K and estimating $A=1-F_\mathrm{spot}/F_\star$ by means of
\texttt{BT-Settl} atmospheric models \citep{allard2011}, the maximum
correction\footnote{i.e.,  when no starspots are occulted during the
  transit: case with $f_\mathrm{occ} =0$ \citep{ballerini2012}} we
  compute is

\begin{equation}
\label{deltak}
\begin{array}{lcl}
\Delta k_{\max} \,(B)/k  & = & -0.0098   \\
\Delta k_{\max} \,(R)/k  & = & -0.0097   \\
\end{array} \qquad (\texttt{T1})\; , \\
\end{equation}
that is, in absolute terms, $\Delta k_{\max} \,(B)=-0.00115$ and
$\Delta k_{\max} \,(R)=-0.00114$.

These corrections can then be applied to the apparent $B$ and $R$
radius ratios $k(B)$, $k(R)$ to obtain lower limits for them, named
$k'(B)$, $k'(R)$ and reported in the last column of Table
\ref{results}.  These corrections might appear rather significant, as
they change $k$ by about 1\%. Nevertheless, the color signature
introduced by  stellar activity is tiny, because the \emph{differential}
correction  $\Delta k_{\max} \,(B) -\Delta k_{\max} \,(R) < 0.0001$ is
well within the measurement error bars $\sigma(k,B)=0.0009$ and
$\sigma(k,R)=0.0004$, respectively. We emphasize that  $\Delta
k_{\max}$ represents a maximal correction  and that the true values
probably lie somewhere between $k$ and $k'$.

By comparing $k'(B)=0.1165\pm0.009$ and $k'(R)=0.1164\pm 0.0004$,  it
is straightforward to conclude that the radius of GJ1214b is not
dependent on wavelength within the error bars, even after the effect
of unocculted starspots has been accounted for. In other words, based
on our LBC data alone, where both the time- and wavelength-dependent
activity effects are corrected for, the continuum of the transmission
spectrum is flat  within $\lesssim$1\% between $B$ and $R$.  In
the next section, we complement our measurements with others
published in the literature through several bandpasses. As for the
latter,  only the apparent values of $k$ are available because
most authors of  previous studies did not apply any correction for
stellar activity.  For consistency, we are forced to adopt our
uncorrected $k$ values throughout the comparison. However, we
demonstrated above that the color effect given by the maximal
correction (computed on the bluest and most affected band of the
whole set: LBC $B$)  is negligible with respect to the measurement
errors,  even during a phase of high stellar activity (2012). We are
therefore confident  that such a correction --even if it were
available-- would change  the mean level of the transmission spectrum,
but not the slope, leaving the main scientific result unchanged.

It is worth noting that the corrections computed above}  neglect the
contribution from  the fraction of spots that are always visible
during a rotation cycle (because they are located in a polar region)
or those that are uniformly distributed over the surface.  The latter case is the
most likely scenario, as fully convective stars at $\mstar \lesssim
0.35~\msun$, such as GJ1214,  are expected to drive activity in a
turbulent dynamo regime rather than with a classic  $\alpha\Omega$
dynamo (\citealt{barnes2011} and references therein).  It is
easy to check what would happen if the filling factor due to the
component of homogeneously distributed spots is the maximum allowed by
previous studies  for an M4V star ($\sim$0.20;
\citealt{berdyugina2005}). Models predict that the correction would
scale approximately as a linear function of the filling factor, that is,
it would be about ten times larger:
\begin{equation}
\label{deltak20}
\begin{array}{lcl}
\Delta k_{\max} \,(B)/k  & = & -0.1173   \\
\Delta k_{\max} \,(R)/k  & = & -0.1150   \\
\end{array} \\
\end{equation}
that is, in absolute terms, $\Delta k_{\max} \,(B)=-0.01378$ and
$\Delta k_{\max} \,(R)=-0.01351$.
We conclude that even after applying such a maximal
correction $\Delta k_{\max} \,(B) -\Delta k_{\max} \,(R) \simeq
0.0003$, the resulting $k'(B)$ and $k'(R)$ would be consistent with
each other within the error bars, leaving the main scientific result
unchanged.

\section{Discussion}\label{discussion}

\begin{figure*}
\centering \includegraphics[width=1.8\columnwidth]{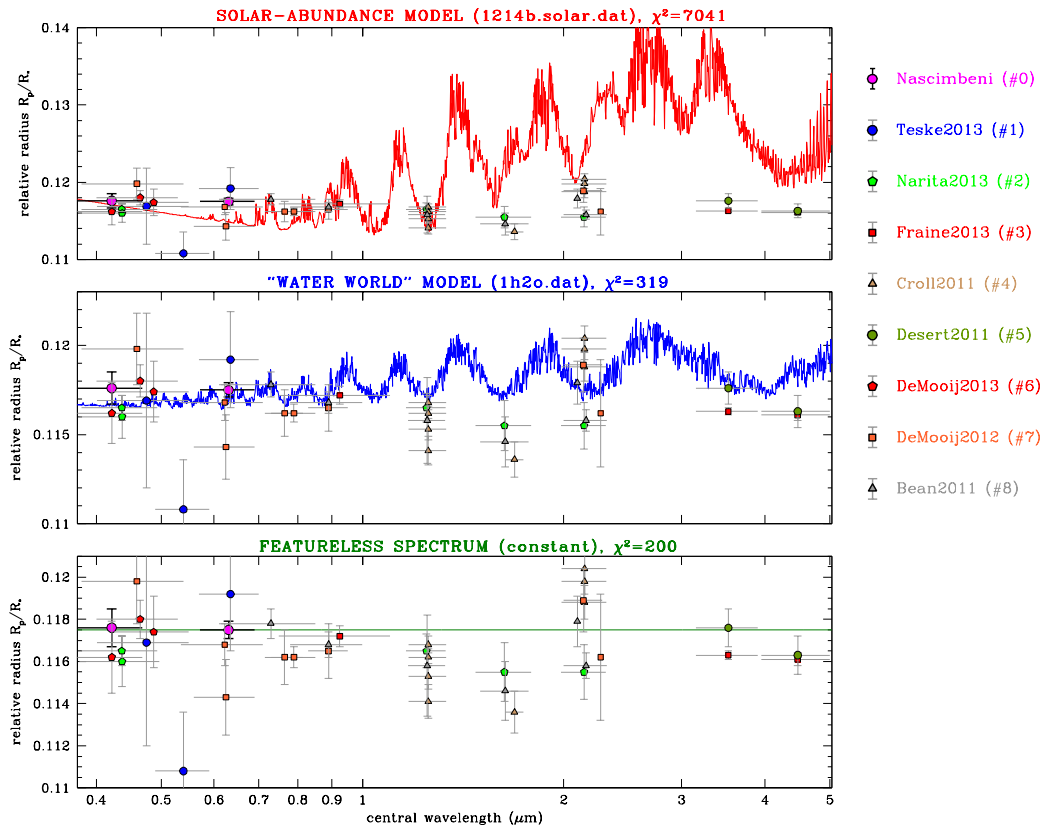}
\caption{Transmission spectrum of GJ1214b reconstructed by merging
  results from the present work (magenta circles) with all the
  published data that assumed the same prior as we did on $i$ and
  $a/\rstar$ (see the legend on the right).  Theoretical models with
  different compositions calculated for GJ1214b by \citet{howe2012}
  are plotted with solid lines. \emph{Upper panel:}  low $\mu$,
  cloud-free atmosphere with solar composition (red
  line). \emph{Middle panel:} high $\mu$, cloud-free atmosphere
  composed by pure water (blue line). \emph{Lower panel:} featureless
  (constant) spectrum simulating a high-altitude layer
  of opaque clouds (green line). The corresponding $\chi^2$ values are
  reported for each case.}
\label{spectra}
\end{figure*}

We reconstructed a transmission spectrum of GJ1214b by merging the two
values of $k$ derived in this work (magenta circles on
Fig. \ref{spectra}) with all the published data points that assumed
the same prior as we did on orbital inclination and scaled  semi-axis
(centered on $i=88^\circ.94$ and $a/\rstar = 14.97$, first adopted by
\citealt{bean2010}).  As explained in Sect.~\ref{TPE}, this is to avoid
a  systematic bias on $k$, because $k$ and $i$ (and, to a much lesser
extent, $a/\rstar$) are known to be strongly  correlated fit
parameters. Unfortunately, this choice excludes some very high
S/N data sets (including \citealt{berta2012} and
\citealt{kreidberg2014}, who fixed $i$ at $89^\circ.30$ and
$i=89^\circ.1$, respectively), but a full homogeneous reanalysis of
these data is beyond the scope of this paper.  The literature data we
gathered include the following light curves, which resulted in 40 data
points after combining measurements gathered through the same
instrument and setup:
\begin{itemize}
\item  one transit from FORS2 at the VLT (spectrophotometry 780--1000~nm; SP)
  by \citet{bean2010}, as later reanalyzed by \citet{bean2011};
\item three transits from FORS2 at the VLT (610--850~nm SP),  MMIRS at the Magellan
  ($H+K$ SP) and HAWKI at the VLT (narrow filter NB2090) by \citet{bean2011}; 
\item two transits from IRAC at the Spitzer (IRAC 3.6~$\mu$m and   4.6~$\mu$m
  bands) by \citet{desert2011};
\item four transits from WIRCam at the CFHT (each light curve was gathered by
  alternatively switching between pair of filters chosen among $J$,
  $K_\mathrm{s}$, and a narrow $\chfour$On), by \citet{croll2011};
\item two transits from WFC at the INT (Sloan $r$ and $I$), one transit from
  NOTCam at the NOT ($K_\mathrm{s}$), one  transit from LIRIS at the WHT
  ($K_\mathrm{c}$), and one transit simultaneously observed through four filters
  (Sloan $g$, $r$, $i$, $z$)  from GROND at the MPI-2.2m, by
  \citet{demooij2012}; 
\item seventeen transits from IRAC at the Spitzer (fourteen at 4.5~$\mu$m and
  three at 3.6~$\mu$m) and seven ground-based transits from TRAPPIST
  (through a wide band encompassing $I+z'$) by \citet{fraine2013};
\item three transits from ACAM at the WHT (Sloan $g$), FORS2 at the VLT (Bessel $B$)
  and WFC at the INT (Sloan $g$) respectively, by \citet{demooij2013}; 
\item four transits from Mont4k at the Kuiper-1.55m (two through Harris $V$
  and two through Harris $R$) and five transits from WiFSIP at STELLA
  (Sloan $g'$), by \citet{teske2013};
\item two transits from SuprimeCam at the Subaru and  FOCAS at the S
  u\-ba\-ru (both in Bessel $B$), and one transit simultaneously observed 
  through $J$, $H$, $K$ from SIRIUS at the IRSF, by \citet{narita2013b}. 
\end{itemize}

Theoretical models of different chemical and physical compositions,
calculated for GJ1214b by \citet{howe2012}, are plotted with solid
lines of different colors in Fig.~\ref{spectra}, along with the
mentioned  data. We focus on three representative cases: 1) a  low
$\mu$, cloud-free atmosphere with solar composition, that is, dominated
by hydrogen and helium (red line); 2) a high $\mu$, cloud-free
atmosphere composed of pure water vapor ($\htwoo$; blue line); 3) a
featureless (constant) spectrum simulating a
high-altitude layer of opaque clouds (green line), which mask every
spectral feature behind that layer.  We integrated the model spectrum
over the photometric passband of each data point to calculate  the
residuals between model and observation, then we performed a
statistical analysis.  The corresponding $\chi^2$ values per degree of
freedom are reported for each case in the corresponding diagram of
Fig.~\ref{spectra}. 

The solar-abundance model is clearly ruled out at high confidence, the
resulting $\chisq$ being 7041.
A similar conclusion was already drawn  by
\citet{bean2010,bean2011} and \citet{desert2011},  was debated by
\citet{croll2011}, and later confirmed by most recent studies.  The
steep increase of absorption at the blue end  of this particular
model spectrum, which is a signature of Rayleigh scattering by
molecular  hydrogen $\htwo$, is ruled out at $3\sigma$ even by the LBC
data alone, which makes ours an independent confirmation. 

The water-based model returns a better fit, although with a rather high
value of $\chisq = 319$, 
while the simpler featureless spectrum yields  
$\chisq = 200$. 
Based
on this comparison, we selected the latter as  the model that best
explains the observations. We emphasize that no other  model
atmosphere among those computed by \citet{howe2012} returns a smaller
$\chisq$, while the best-fit $\chisqr$ found above implies 
error bars underestimated by a factor of $\sim$2, under the hypothesis
of an underlying Gaussian distribution. This is frequently
observed when comparing results originating from different studies
(\citealt{southworth2011}, among others, is  a rich source of such
examples). Possible reasons include the use of poorly suited numerical
algorithms to estimate the uncertainties over the parameters, which
cannot be calculated without taking into account the presence of
highly correlated fit variables. A second explanation is stellar
variability, which can alter most transit parameters (including $k$,
the most relevant one to our purposes) as a function of both time and
wavelength.  For GJ1214b, we expect variability to be the
dominant source of scatter in its optical transmission spectrum,
following the discussion made in Sect.~\ref{kmax}, where we estimated
that the peak-to-valley relative variation of $k$ owing to stellar
activity could reach a few percent under reasonable assumptions. 

It is more challenging, however, to explain why the apparent radius of
GJ1214b  changes so much when measured at different epochs in the NIR,
where  spot contrast is expected to be lower and the corresponding
effect  on the apparent $k$ weaker.  This is most strikingly
noticeable in the $K$ band, where different authors found radii
discrepant by more than 4$\sigma$ \citep{narita2013b}. On one hand,
little is known about the detailed magnetic behavior of stars close to
the M5V transition where GJ1214 lies, and the $\teff$
vs.~$T_\mathrm{spot}$ relation reported by \citet{berdyugina2005} stops at
$\teff = 3300$~K. It has been hypothesized that much colder spots
could begin to appear at those  spectral types
\citep{barnes2011}. That would explain why many  surveys of mid- and
late-M stars measured a photometric jitter that is similar to
that usually reported in the optical region (\citealt{plavchan2008},
among others).
 
On the other hand, another plausible interpretation for these cases is
the presence of residual red noise, which is known to plague NIR data,
often in subtle ways \citep{swain2013}.

\section{Conclusions}\label{conclusions}

We demonstrated that the dual-channel capabilities of  the LBC
prime-focus camera and its unrivaled efficiency  (especially on the
blue side of the spectrum) are extremely effective in delivering
high-precision  photometric time series of faint, red, and active
targets such as GJ1214b.  Indeed, the Bessel $B$ and $R$ light curves
presented in this work are the most precise ever published in their
respective  spectral regions. Not less important, they are truly
simultaneous measurements, which enabled us to remove the
time-dependent component of stellar activity. We exploited a smaller, robotic
facility (WiFSIP at STELLA) to monitor the photometric variability of
GJ1214 over a larger time scale to determine the wavelength-dependent 
contribution. This resulted in a new, robust
estimate of the stellar rotational  period at $\prot\simeq 80$~d. By
comparing the WiFSIP light curves with those reported in the literature, we also
noted that the activity  level of GJ1214 was significantly higher
than the average  throughout the 2012 observing season. The detection
of a large bump in the second transit (\texttt{T2}; 2012/05/17),
induced by the crossing of a starspot, appears to support that
finding. The nature of this bump is unambiguously confirmed by its
$B-R$ signature. The latter allowed us to derive a lower limit to the
temperature contrast between the  photosphere and the spot ($\Delta T
\simeq 110$~K), again highlighting the advantages of multicolor
photometry.

The relative radii of GJ1214b through the $B$ and $R$ filter from the
LBC data alone, once corrected for the presence of unocculted spots,
are perfectly consistent with each  other within $\sim 1\%$. This
finding suggests a flat transmission spectrum in the optical region and
rules out at $5\sigma$ the tentative detection of Rayleigh scattering
by $\htwo$ that has been claimed by some previous works
\citep{demooij2012,fraine2013}. After combining our points with other
optical and NIR measurements, we rule out cloud-free atmospheric
models dominated by  H/He and by heavier compositions such as pure
water vapor.  We conclude that a flat transmission spectrum, such as
that  resulting from an atmosphere with a thick cloud layer  at high
altitude, is still the model that best explains the available data.
Our conclusion is consistent with other recent independent works (most
notably, \citealt{kreidberg2014}) that also ruled out cloud-free,
high-$\mu$ models in favor of a featureless  spectrum. We have to
await the commissioning of future space-based  instruments (such as the
James Webb Space Telescope) to further investigate the atmosphere of
GJ1214 and confirm the cloudy scenario \citep{belu2011}.

\begin{acknowledgements}

\emph{We acknowledge the support from the LBT-Italian Coordination Facility for the
execution of observations, data distribution and reduction.
V.~N.~and G.~P.~acknowledge partial support by the
Universit\`a di Padova through the ``progetto di Ateneo \#CPDA103591''.
V.~N.~acknowledges partial support from INAF-OAPd through
the grant ``Analysis of HARPS-N data in the framework of GAPS project''
(\#19/2013) and ``Studio preparatorio per le osservazioni della 
missione ESA/CHEOPS'' (\#42/2013).
We thank Thomas Granzer for his extensive technical support with 
the STELLA monitoring campaign. 
Some tasks of our data analysis have been carried out with
the VARTOOLS \citep{hartman2008} and \texttt{Astrometry.net} codes
\citep{lang2010}. This research has made use of the International 
Variable Star Index (VSX) database, operated at AAVSO, Cambridge, 
Massachusetts, USA.} VN dedicates this paper to the memory of 
Renzo Nascimbeni (1942--2014).

\end{acknowledgements}

\bibliographystyle{aa}
\bibliography{biblio}

\begin{thebibliography}{60}
\expandafter\ifx\csname natexlab\endcsname\relax\def\natexlab#1{#1}\fi

\bibitem[{{Afram} \& {Berdyugina}(2015)}]{afram2015}
{Afram}, N. \& {Berdyugina}, S.~V. 2015, \aap, 576, A34

\bibitem[{{Allard} {et~al.}(2011){Allard}, {Homeier}, \&
  {Freytag}}]{allard2011}
{Allard}, F., {Homeier}, D., \& {Freytag}, B. 2011, in Astronomical Society of
  the Pacific Conference Series, Vol. 448, 16th Cambridge Workshop on Cool
  Stars, Stellar Systems, and the Sun, ed. C.~{Johns-Krull}, M.~K. {Browning},
  \& A.~A. {West}, 91

\bibitem[{{Anderson} \& {King}(1999)}]{ak1999}
{Anderson}, J. \& {King}, I.~R. 1999, \pasp, 111, 1095

\bibitem[{{Anglada-Escud{\'e}} {et~al.}(2013){Anglada-Escud{\'e}},
  {Rojas-Ayala}, {Boss}, {Weinberger}, \& {Lloyd}}]{anglada2013}
{Anglada-Escud{\'e}}, G., {Rojas-Ayala}, B., {Boss}, A.~P., {Weinberger},
  A.~J., \& {Lloyd}, J.~P. 2013, \aap, 551, A48

\bibitem[{{Ballerini} {et~al.}(2012){Ballerini}, {Micela}, {Lanza}, \&
  {Pagano}}]{ballerini2012}
{Ballerini}, P., {Micela}, G., {Lanza}, A.~F., \& {Pagano}, I. 2012, \aap, 539,
  A140

\bibitem[{{Barnes} {et~al.}(2011){Barnes}, {Jeffers}, \& {Jones}}]{barnes2011}
{Barnes}, J.~R., {Jeffers}, S.~V., \& {Jones}, H.~R.~A. 2011, \mnras, 412, 1599

\bibitem[{{Barstow} {et~al.}(2013){Barstow}, {Aigrain}, {Irwin}, {Fletcher}, \&
  {Lee}}]{barstow2013}
{Barstow}, J.~K., {Aigrain}, S., {Irwin}, P.~G.~J., {Fletcher}, L.~N., \&
  {Lee}, J.-M. 2013, \mnras, 434, 2616

\bibitem[{{Bean} {et~al.}(2011){Bean}, {D{\'e}sert}, {Kabath}, {Stalder},
  {Seager}, {Miller-Ricci Kempton}, {Berta}, {Homeier}, {Walsh}, \&
  {Seifahrt}}]{bean2011}
{Bean}, J.~L., {D{\'e}sert}, J.-M., {Kabath}, P., {et~al.} 2011, \apj, 743, 92

\bibitem[{{Bean} {et~al.}(2010){Bean}, {Miller-Ricci Kempton}, \&
  {Homeier}}]{bean2010}
{Bean}, J.~L., {Miller-Ricci Kempton}, E., \& {Homeier}, D. 2010, \nat, 468,
  669

\bibitem[{{Belu} {et~al.}(2011){Belu}, {Selsis}, {Morales}, {Ribas}, {Cossou},
  \& {Rauer}}]{belu2011}
{Belu}, A.~R., {Selsis}, F., {Morales}, J.-C., {et~al.} 2011, \aap, 525, A83

\bibitem[{{Berdyugina}(2005)}]{berdyugina2005}
{Berdyugina}, S.~V. 2005, Living Reviews in Solar Physics, 2, 8

\bibitem[{{Berta} {et~al.}(2011){Berta}, {Charbonneau}, {Bean}, {Irwin},
  {Burke}, {D{\'e}sert}, {Nutzman}, \& {Falco}}]{berta2011}
{Berta}, Z.~K., {Charbonneau}, D., {Bean}, J., {et~al.} 2011, \apj, 736, 12

\bibitem[{{Berta} {et~al.}(2012){Berta}, {Charbonneau}, {D{\'e}sert},
  {Miller-Ricci Kempton}, {McCullough}, {Burke}, {Fortney}, {Irwin}, {Nutzman},
  \& {Homeier}}]{berta2012}
{Berta}, Z.~K., {Charbonneau}, D., {D{\'e}sert}, J.-M., {et~al.} 2012, \apj,
  747, 35

\bibitem[{{Bertin} \& {Arnouts}(1996)}]{bertin1996}
{Bertin}, E. \& {Arnouts}, S. 1996, \aaps, 117, 393

\bibitem[{{Carter} \& {Winn}(2009)}]{carter2009}
{Carter}, J.~A. \& {Winn}, J.~N. 2009, \apj, 704, 51

\bibitem[{{Carter} {et~al.}(2011){Carter}, {Winn}, {Holman}, {Fabrycky},
  {Berta}, {Burke}, \& {Nutzman}}]{carter2011}
{Carter}, J.~A., {Winn}, J.~N., {Holman}, M.~J., {et~al.} 2011, \apj, 730, 82

\bibitem[{{Charbonneau} {et~al.}(2009){Charbonneau}, {Berta}, {Irwin}, {Burke},
  {Nutzman}, {Buchhave}, {Lovis}, {Bonfils}, {Latham}, {Udry}, {Murray-Clay},
  {Holman}, {Falco}, {Winn}, {Queloz}, {Pepe}, {Mayor}, {Delfosse}, \&
  {Forveille}}]{charbonneau2009}
{Charbonneau}, D., {Berta}, Z.~K., {Irwin}, J., {et~al.} 2009, \nat, 462, 891

\bibitem[{{Charbonneau} {et~al.}(2002){Charbonneau}, {Brown}, {Noyes}, \&
  {Gilliland}}]{charbonneau2002}
{Charbonneau}, D., {Brown}, T.~M., {Noyes}, R.~W., \& {Gilliland}, R.~L. 2002,
  \apj, 568, 377

\bibitem[{{Claret}(2004)}]{claret2004}
{Claret}, A. 2004, \aap, 428, 1001

\bibitem[{{Claret} {et~al.}(2012){Claret}, {Hauschildt}, \&
  {Witte}}]{claret2012}
{Claret}, A., {Hauschildt}, P.~H., \& {Witte}, S. 2012, \aap, 546, A14

\bibitem[{{Croll} {et~al.}(2011){Croll}, {Albert}, {Jayawardhana},
  {Miller-Ricci Kempton}, {Fortney}, {Murray}, \& {Neilson}}]{croll2011}
{Croll}, B., {Albert}, L., {Jayawardhana}, R., {et~al.} 2011, \apj, 736, 78

\bibitem[{{de Mooij} {et~al.}(2012){de Mooij}, {Brogi}, {de Kok},
  {Koppenhoefer}, {Nefs}, {Snellen}, {Greiner}, {Hanse}, {Heinsbroek}, {Lee},
  \& {van der Werf}}]{demooij2012}
{de Mooij}, E.~J.~W., {Brogi}, M., {de Kok}, R.~J., {et~al.} 2012, \aap, 538,
  A46

\bibitem[{{de Mooij} {et~al.}(2013){de Mooij}, {Brogi}, {de Kok}, {Snellen},
  {Croll}, {Jayawardhana}, {Hoekstra}, {Otten}, {Bekkers}, {Haffert}, \& {van
  Houdt}}]{demooij2013}
{de Mooij}, E.~J.~W., {Brogi}, M., {de Kok}, R.~J., {et~al.} 2013, \apj, 771,
  109

\bibitem[{{D{\'e}sert} {et~al.}(2011){D{\'e}sert}, {Bean}, {Miller-Ricci
  Kempton}, {Berta}, {Charbonneau}, {Irwin}, {Fortney}, {Burke}, \&
  {Nutzman}}]{desert2011}
{D{\'e}sert}, J.-M., {Bean}, J., {Miller-Ricci Kempton}, E., {et~al.} 2011,
  \apjl, 731, L40

\bibitem[{{Fraine} {et~al.}(2013){Fraine}, {Deming}, {Gillon}, {Jehin},
  {Demory}, {Benneke}, {Seager}, {Lewis}, {Knutson}, \&
  {D{\'e}sert}}]{fraine2013}
{Fraine}, J.~D., {Deming}, D., {Gillon}, M., {et~al.} 2013, \apj, 765, 127

\bibitem[{{Giallongo} {et~al.}(2008){Giallongo}, {Ragazzoni}, {Grazian},
  {Baruffolo}, {Beccari}, {de Santis}, {Diolaiti}, {di Paola}, {Farinato},
  {Fontana}, {Gallozzi}, {Gasparo}, {Gentile}, {Green}, {Hill}, {Kuhn},
  {Pasian}, {Pedichini}, {Radovich}, {Salinari}, {Smareglia}, {Speziali},
  {Testa}, {Thompson}, {Vernet}, \& {Wagner}}]{giallongo2008}
{Giallongo}, E., {Ragazzoni}, R., {Grazian}, A., {et~al.} 2008, \aap, 482, 349

\bibitem[{{Harps{\o}e} {et~al.}(2013){Harps{\o}e}, {Hardis}, {Hinse},
  {J{\o}rgensen}, {Mancini}, {Southworth}, {Alsubai}, {Bozza}, {Browne},
  {Burgdorf}, {Calchi Novati}, {Dodds}, {Dominik}, {Fang}, {Finet}, {Gerner},
  {Gu}, {Hundertmark}, {Jessen-Hansen}, {Kains}, {Kerins}, {Kjeldsen},
  {Liebig}, {Lund}, {Lundkvist}, {Mathiasen}, {Nesvorn{\'y}}, {Nikolov},
  {Penny}, {Proft}, {Rahvar}, {Ricci}, {Sahu}, {Scarpetta}, {Sch{\"a}fer},
  {Sch{\"o}nebeck}, {Snodgrass}, {Skottfelt}, {Surdej}, {Tregloan-Reed}, \&
  {Wertz}}]{harpsoe2013}
{Harps{\o}e}, K.~B.~W., {Hardis}, S., {Hinse}, T.~C., {et~al.} 2013, \aap, 549,
  A10

\bibitem[{{Hartman} {et~al.}(2008){Hartman}, {Gaudi}, {Holman}, {McLeod},
  {Stanek}, {Barranco}, {Pinsonneault}, \& {Kalirai}}]{hartman2008}
{Hartman}, J.~D., {Gaudi}, B.~S., {Holman}, M.~J., {et~al.} 2008, \apj, 675,
  1254

\bibitem[{{Howe} \& {Burrows}(2012)}]{howe2012}
{Howe}, A.~R. \& {Burrows}, A.~S. 2012, \apj, 756, 176

\bibitem[{{Howell}(2006)}]{howell2006}
{Howell}, S.~B. 2006, {Handbook of CCD astronomy}, ed. R.~{Ellis}, J.~{Huchra},
  S.~{Kahn}, G.~{Rieke}, \& P.~B. {Stetson}

\bibitem[{{Jackson} \& {Jeffries}(2013)}]{jackson2013}
{Jackson}, R.~J. \& {Jeffries}, R.~D. 2013, \mnras, 431, 1883

\bibitem[{{Kipping}(2010)}]{kipping2010b}
{Kipping}, D.~M. 2010, \mnras, 407, 301

\bibitem[{{Kreidberg} {et~al.}(2014){Kreidberg}, {Bean}, {D{\'e}sert},
  {Benneke}, {Deming}, {Stevenson}, {Seager}, {Berta-Thompson}, {Seifahrt}, \&
  {Homeier}}]{kreidberg2014}
{Kreidberg}, L., {Bean}, J.~L., {D{\'e}sert}, J.-M., {et~al.} 2014, \nat, 505,
  69

\bibitem[{{Kundurthy} {et~al.}(2011){Kundurthy}, {Agol}, {Becker}, {Barnes},
  {Williams}, \& {Mukadam}}]{kundurthy2011}
{Kundurthy}, P., {Agol}, E., {Becker}, A.~C., {et~al.} 2011, \apj, 731, 123

\bibitem[{{Lang} {et~al.}(2010){Lang}, {Hogg}, {Mierle}, {Blanton}, \&
  {Roweis}}]{lang2010}
{Lang}, D., {Hogg}, D.~W., {Mierle}, K., {Blanton}, M., \& {Roweis}, S. 2010,
  \aj, 139, 1782

\bibitem[{{Morley} {et~al.}(2013){Morley}, {Fortney}, {Kempton}, {Marley},
  {Visscher}, \& {Zahnle}}]{morley2013}
{Morley}, C.~V., {Fortney}, J.~J., {Kempton}, E.~M.-R., {et~al.} 2013, \apj,
  775, 33

\bibitem[{{Murgas} {et~al.}(2012){Murgas}, {Pall{\'e}}, {Cabrera-Lavers},
  {Col{\'o}n}, {Mart{\'{\i}}n}, \& {Parviainen}}]{murgas2012}
{Murgas}, F., {Pall{\'e}}, E., {Cabrera-Lavers}, A., {et~al.} 2012, \aap, 544,
  A41

\bibitem[{{Narita} {et~al.}(2013){Narita}, {Fukui}, {Ikoma}, {Hori},
  {Kurosaki}, {Kawashima}, {Nagayama}, {Onitsuka}, {Sukom}, {Nakajima},
  {Tamura}, {Kuroda}, {Yanagisawa}, {Hirano}, {Kawauchi}, {Kuzuhara}, {Ohnuki},
  {Suenaga}, {Takahashi}, {Izumiura}, {Kawai}, \& {Yoshida}}]{narita2013b}
{Narita}, N., {Fukui}, A., {Ikoma}, M., {et~al.} 2013, \apj, 773, 144

\bibitem[{{Nascimbeni} {et~al.}(2013{\natexlab{a}}){Nascimbeni}, {Cunial},
  {Murabito}, {Sada}, {Aparicio}, {Piotto}, {Bedin}, {Milone}, {Rosenberg},
  {Zurlo}, {Borsato}, {Damasso}, {Granata}, \& {Malavolta}}]{nascimbeni2013a}
{Nascimbeni}, V., {Cunial}, A., {Murabito}, S., {et~al.} 2013{\natexlab{a}},
  \aap, 549, A30

\bibitem[{{Nascimbeni} {et~al.}(2011){Nascimbeni}, {Piotto}, {Bedin}, \&
  {Damasso}}]{nascimbeni2011a}
{Nascimbeni}, V., {Piotto}, G., {Bedin}, L.~R., \& {Damasso}, M. 2011, \aap,
  527, A85+

\bibitem[{{Nascimbeni} {et~al.}(2013{\natexlab{b}}){Nascimbeni}, {Piotto},
  {Pagano}, {Scandariato}, {Sani}, \& {Fumana}}]{nascimbeni2013b}
{Nascimbeni}, V., {Piotto}, G., {Pagano}, I., {et~al.} 2013{\natexlab{b}},
  \aap, 559, A32

\bibitem[{{Oshagh} {et~al.}(2013){Oshagh}, {Santos}, {Boisse}, {Bou{\'e}},
  {Montalto}, {Dumusque}, \& {Haghighipour}}]{oshagh2013}
{Oshagh}, M., {Santos}, N.~C., {Boisse}, I., {et~al.} 2013, \aap, 556, A19

\bibitem[{{Plavchan} {et~al.}(2008){Plavchan}, {Jura}, {Kirkpatrick}, {Cutri},
  \& {Gallagher}}]{plavchan2008}
{Plavchan}, P., {Jura}, M., {Kirkpatrick}, J.~D., {Cutri}, R.~M., \&
  {Gallagher}, S.~C. 2008, \apjs, 175, 191

\bibitem[{{Pont} {et~al.}(2007){Pont}, {Gilliland}, {Moutou}, {Charbonneau},
  {Bouchy}, {Brown}, {Mayor}, {Queloz}, {Santos}, \& {Udry}}]{pont2007}
{Pont}, F., {Gilliland}, R.~L., {Moutou}, C., {et~al.} 2007, \aap, 476, 1347

\bibitem[{{Redfield} {et~al.}(2008){Redfield}, {Endl}, {Cochran}, \&
  {Koesterke}}]{redfield2008}
{Redfield}, S., {Endl}, M., {Cochran}, W.~D., \& {Koesterke}, L. 2008, \apjl,
  673, L87

\bibitem[{{Scandariato} \& {Micela}(2014)}]{scandariato2014}
{Scandariato}, G. \& {Micela}, G. 2014, Experimental Astronomy

\bibitem[{{Seager} \& {Deming}(2010)}]{seager2010}
{Seager}, S. \& {Deming}, D. 2010, \araa, 48, 631

\bibitem[{{Sing} {et~al.}(2011){Sing}, {Pont}, {Aigrain}, {Charbonneau},
  {D{\'e}sert}, {Gibson}, {Gilliland}, {Hayek}, {Henry}, {Knutson}, {Lecavelier
  Des Etangs}, {Mazeh}, \& {Shporer}}]{sing2011}
{Sing}, D.~K., {Pont}, F., {Aigrain}, S., {et~al.} 2011, \mnras, 416, 1443

\bibitem[{{Southworth}(2008)}]{southworth2008}
{Southworth}, J. 2008, \mnras, 386, 1644

\bibitem[{{Southworth}(2011)}]{southworth2011}
{Southworth}, J. 2011, \mnras, 417, 2166

\bibitem[{{Southworth} {et~al.}(2004){Southworth}, {Maxted}, \&
  {Smalley}}]{southworth2004}
{Southworth}, J., {Maxted}, P.~F.~L., \& {Smalley}, B. 2004, \mnras, 351, 1277

\bibitem[{{Southworth} {et~al.}(2005){Southworth}, {Smalley}, {Maxted},
  {Claret}, \& {Etzel}}]{southworth2005}
{Southworth}, J., {Smalley}, B., {Maxted}, P.~F.~L., {Claret}, A., \& {Etzel},
  P.~B. 2005, \mnras, 363, 529

\bibitem[{{Strassmeier}(2009)}]{strassmeier2009}
{Strassmeier}, K.~G. 2009, \aapr, 17, 251

\bibitem[{{Strassmeier} {et~al.}(2004){Strassmeier}, {Granzer}, {Weber},
  {Woche}, {Andersen}, {Bartus}, {Bauer}, {Dionies}, {Popow}, {Fechner},
  {Hildebrandt}, {Washuettl}, {Ritter}, {Schwope}, {Staude}, {Paschke},
  {Stolz}, {Serre-Ricart}, {de la Rosa}, \& {Arnay}}]{strassmeier2004}
{Strassmeier}, K.~G., {Granzer}, T., {Weber}, M., {et~al.} 2004, Astronomische
  Nachrichten, 325, 527

\bibitem[{{Swain} {et~al.}(2013){Swain}, {Deroo}, \& {Wagstaff}}]{swain2013}
{Swain}, M.~R., {Deroo}, P., \& {Wagstaff}, K.~L. 2013, ArXiv e-prints

\bibitem[{{Teske} {et~al.}(2013){Teske}, {Turner}, {Mueller}, \&
  {Griffith}}]{teske2013}
{Teske}, J.~K., {Turner}, J.~D., {Mueller}, M., \& {Griffith}, C.~A. 2013,
  \mnras, 431, 1669

\bibitem[{{Valencia} {et~al.}(2013){Valencia}, {Guillot}, {Parmentier}, \&
  {Freedman}}]{valencia2013}
{Valencia}, D., {Guillot}, T., {Parmentier}, V., \& {Freedman}, R.~S. 2013,
  \apj, 775, 10

\bibitem[{{Walkowicz} {et~al.}(2013){Walkowicz}, {Basri}, \&
  {Valenti}}]{walkowicz2013}
{Walkowicz}, L.~M., {Basri}, G., \& {Valenti}, J.~A. 2013, \apjs, 205, 17

\bibitem[{{Weber} {et~al.}(2012){Weber}, {Granzer}, \&
  {Strassmeier}}]{weber2012}
{Weber}, M., {Granzer}, T., \& {Strassmeier}, K.~G. 2012, in Society of
  Photo-Optical Instrumentation Engineers (SPIE) Conference Series, Vol. 8451,
  Society of Photo-Optical Instrumentation Engineers (SPIE) Conference Series

\bibitem[{{Zechmeister} \& {K{\"u}rster}(2009)}]{zechmeister2009}
{Zechmeister}, M. \& {K{\"u}rster}, M. 2009, \aap, 496, 577

\end{thebibliography}

\end{document}